\documentclass[]{aa}
\usepackage[T1]{fontenc}
\usepackage[utf8]{inputenc}

\usepackage{graphicx}
\usepackage{natbib}
\usepackage{ulem}
\usepackage{textcomp}
\usepackage{gensymb}
\usepackage{longtable}
\usepackage{threeparttable}
\usepackage{multicol}
\usepackage{multirow}
\usepackage{float}
\usepackage{todonotes}
\usepackage[Symbol]{upgreek}
\usepackage{array}
\usepackage{amsmath}
\usepackage{amssymb}
\usepackage{etoolbox}
\usepackage{txfonts}
\usepackage{url}
\usepackage{xcolor}

\usepackage[most]{tcolorbox}
\usepackage{hyperref}
\hypersetup{
     colorlinks=true,
     linkcolor=blue,
     filecolor=blue,
     citecolor = blue,      
     urlcolor=blue,
     }
\usepackage{xcolor}

\newcommand{\erosita}{{\small eROSITA}}
\newcommand{\apec}{{\small APEC}}
\newcommand{\healpix}{{\small HEALPix}}
\newcommand{\xspec}{{\small XSPEC}}
\newcommand{\myname}{Goat Horn complex}
\newcommand{\os}{O {\tiny VII}}
\newcommand{\oo}{O {\tiny VIII}}

\begin{document}

\title{Discovery of the \myname{}: a $\sim$1000 deg$^2$ diffuse X-ray source connected to radio loop XII}

\author{
Nicola Locatelli\inst{1,2} \thanks{nicola.locatelli@inaf.it},
Gabriele Ponti\inst{2,1},
Andrea Merloni\inst{1},
Xueying Zheng\inst{1},
Konrad Dennerl\inst{1},
Frank Haberl\inst{1},
Chandreyee Maitra\inst{1},
Jeremy Sanders\inst{1},
Manami Sasaki\inst{3},
Heshou Zhang\inst{2} 
}

\institute{
Max-Planck-Institut f\"ur Extraterrestrische Physik (MPE), Giessenbachstrasse 1, 85748 Garching bei M\"unchen, Germany
\and INAF - Osservatorio Astronomico di Brera, via E. Bianchi 46, 23807 Merate (LC), Italy
\and Dr. Karl Remeis Observatory, Erlangen Centre for Astroparticle Physics, Friedrich-Alexander-Universit\"at Erlangen-N\"urnberg, Sternwartstra{\ss}e 7, 96049 Bamberg, Germany
}

\authorrunning{N. Locatelli et al.}
\titlerunning{The \myname{} and radio loop XII}

\date{Accepted ???. Received ???; in original form ???}

\abstract{
A dozen of patches of polarized radio emission spanning tens of degrees in the form of coherent and stationary loops are observed at radio frequencies across the sky. Their origin is usually associated to nearby shocks, possibly arising from close supernovae explosions. The origin of the radio Loop XII remains so far unknown. We report an anti-correlation of the radio polarized emission of loop XII with a large patch of soft X-ray emission found with SRG/\erosita{} in excess of the background surface brightness, in the same region. The soft X-ray seemingly coherent patch in excess of the background emission, which we dub as the \myname{}, extends over a remarkable area of $\sim 1000$ deg$^2$ and includes an arc-shaped enhancement potentially tracing a cold front. An anti-correlation of the X-ray intensity with the temperature of the plasma responsible for the X-ray emission is also observed. The X-ray bright arc seems to anticipate the radio loop XII by some degrees on the sky. This behaviour can be recast in terms of a correlation between X-ray surface brightness and radio depolarization. We explore and discuss different possible scenarios for the source of the diffuse emission in the Goat Horn complex: a large supernova remnant; an outflow from active star formation regions in nearby Galactic spiral arms; a hot atmosphere around the Large Magellanic Cloud. In order to probe these scenarios further, a more detailed characterization on the velocity of the hot gas is required.
}
\maketitle

\label{firstpage}

\section{Introduction}

Since the advent of the ROSAT All-Sky Survey (RASS, \citealt{1990Ap&SS.171..207S}), the X-ray sky is known to include different extended sources of diffuse X-ray emission, with the North Polar Spur (NPS), the Eridanus-Orion Superbubble, the Monogem, Antlia and Vela Supernova Remnants (SNRs) holding the largest angular size \citep{Zheng2024}. Although these different sources encompass a moderate range in surface brightness from faint (e.g. Monogem, Antlia) to very bright (e.g. NPS), they are the easiest to recognize and distinguish from the background emission by a simple visual inspection of any X-ray soft band map. 
The recent advent of the first eROSITA All-Sky Survey (eRASS1, \citealt{Merloni24}) improved on this view thanks to its higher signal-to-noise ratio (S/N) provided. The quality of the eRASS data has been assessed through a detailed comparison with the RASS data and their consistency has been validated \citep{Zheng2024}. The smaller amount of systematic sources of noise, for example, recently allowed for the recognition of two very large and extended sources of diffuse emission symmetrically displaced around the Galactic center and Galactic plane, known as the eROSITA bubbles \citep{2020Natur.588..227P}. In a similar fashion, in this work we study an additional extended feature found in the eRASS1 data which had not been recognized in the ROSAT maps due to the larger presence of systematics close to the South Ecliptic Pole (SEP). In fact, around the SEP, the sensitivity of both the RASS and eRASS1 data is enhanced by the frequent passage of the telescopes Field of View (FoV) due to the similar survey strategy of both instruments. Although the advantage carried by a higher sensitivity, frequent observations can carry a higher bias in case of a background noise extended in time, such as the long-term enhancement experienced by ROSAT and attributed to a higher solar activity at the time of the RASS, as well as to the passage of ROSAT through the South Atlantic Anomaly, a region of highly increased particle background \citep{1998LNP...506..113F, 2003JGRA..108.8031R}. 
Therefore, the new eRASS1 data potentially allow to detect extended diffuse emission in excess of the background in all the regions affected by systematics in RASS, as happened with the eROSITA bubbles \citep{Zheng2024}. The new coherent patch of diffuse X-ray emission presented in this work remarkably extends over $\rm \sim 1000\,deg^2$ and encompasses several constellations (e.g. Dorado, Mensa, Hydra, Reticulus, Volans, Pictor, Chamaleon). One of its brightest features is an arc-shaped brightening spanning $\geq 10$ deg and resembling in shape a goat's horn. To easily refer to the extended coherent patch of diffuse emission studied in this work (and not easy to associate with a particular constellation), we then dub it as the \myname{}.

In general, multiwavelength observations of diffuse sources and possible correlations between bands are able to provide insight on the nature of the diffuse source. In particular, the co-detection of diffuse X-ray and radio emission usually show the presence of shocks (traced by the synchrotron emission that they induce) heating plasma to very high temperatures (traced by the X-ray thermal emission, with $kT\sim 0.1-1$ keV, \citealt[e.g.,][]{2019Natur.567..347P, 2021A&A...646A..66P}).
Synchrotron emission is often observed by radio surveys at $\sim$GHz frequencies as a major diffuse component \citep[e.g.,][]{2019MNRAS.489.2330C}. At the largest angular scales, the synchrotron emission is usually associated with a Galactic origin, arising by the combination of a Galactic cosmic ray spectrum and the presence of a Galactic magnetic field. Polarized intensity in particular highlights the non-thermal nature (i.e. synchrotron) of the emission and has been used to detect large-scale (i.e. 10-100 deg) circular rings referred to as "radio loops" (\citealt[][and references therein]{2015MNRAS.452..656V}; the loops are numbered and named  after their projected length). These loops are usually thought to be linked to local Galactic phenomena in general, auch as nearby supernova remnants, Galactic outflows or associated to OB stars complexes \citep{2023A&A...677L..11B}.

In Sec.~\ref{sec:obs} we present the different datasets used in this work; in Sec.~\ref{sec:results} we present the observational evidences of the presence of an additional source of extended emission, never recognized before; in Sec.~\ref{sec:discussion} we link the observational evidences to test different hypotheses on the nature of the new extended source; in Sec~\ref{sec:conclusion} we derive our conclusions and open questions.

\section{Data} \label{sec:obs}

\subsection{X-ray intensity}
\begin{figure*}
    \centering
    \includegraphics[width=0.49\textwidth]{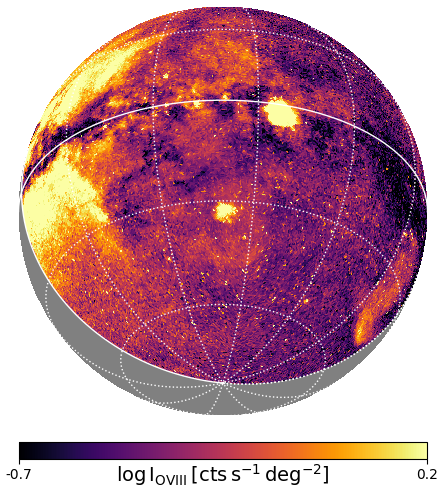}
    \includegraphics[width=0.49\textwidth]{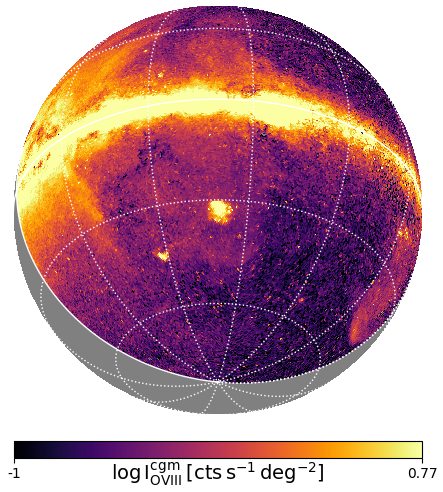}
    \caption{\oo{} narrowband eRASS1 maps. Left panel: The image is centered at the LMC coordinates $(l,b)=(280,-33.7)$ deg. The same projection is kept for all the maps presented in this work. The solid white line represent the great circles $b=0$ deg and $l=0$ deg. The dotted white lines are separated by $\Delta l=30$ deg in longitude and $\Delta b=30$ deg in latitude.
    Right panel: deabsorbed \oo{} eRASS1 map of the warm-hot CGM component of the Milky Way (see text for details on the deabsorption method).
    }
    \label{fig:hmap_O8cgm_n128}
\end{figure*}
\begin{figure}
    \centering
    \includegraphics[width=0.49\textwidth]{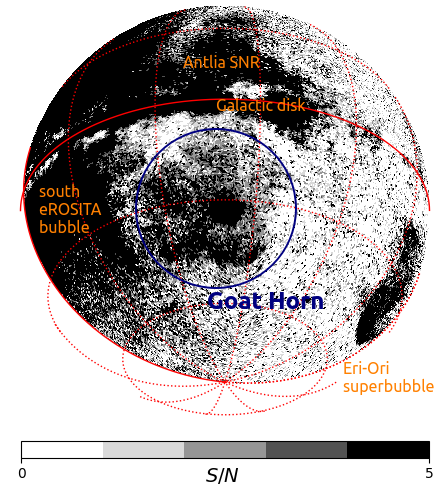}
    \caption{$S/N$ map of the \oo{} deabsorbed image. $S/N$ is the RMS value of the deabsorbed \oo{} map at $220<l<250$ deg and fixed $b$ ($\pm 5$ deg). The longitude range has been selected to represent the soft X-ray background intensity. Coherent patches at $S/N\geq 2$ are associated with discrete extended sources. The new Goat Horn complex is defined by the coherent $S/N\geq 2$ patch approximately within the blue circle.}
    \label{fig:hmap_sigma}
\end{figure}

The X-ray observations have been carried by the extended ROentgen Survey with an Imaging Telescope Array (\erosita, \citealt{2021A&A...647A...1P}) onboard the SRG observatory \citep{2021A&A...656A.132S}, during eRASS1. The data selection, analysis and imaging are the same as described in \cite[][we refer the reader there for further details]{Zheng2024} and in Zheng et al., (submitted). 
In Fig.~\ref{fig:hmap_O8cgm_n128} we show a narrowband image around the bright soft X-ray \oo{} line (0.614-0.694 keV) in a region around the polarized radio loop XII \citep{2015MNRAS.452..656V} and centered on the position of the Large Magellanic Cloud (LMC).

By selecting photons in narrow energy bands centered on the energy of bright emission lines, Zheng et al. (submitted) have built maps of the two high-ionization states of the oxygen (i.e. \os,\oo), thought to trace the warm-hot phase ($T\sim 10^6$ K) of the Milky Way circumgalactic medium \citep[e.g.][]{2023arXiv231010715L}. 
To detect diffuse emission in excess of the background, we define as background region the stripe encompassing the longitude range $250<l<220$ deg. This stripe encompasses all latitudes $b \in [-90;90]$ deg. The longitude range, has been selected by the absence of evident structures (see Fig.~\ref{fig:hmap_O8cgm_n128}) in the X-ray background and is considered as representative of the emission produced by the warm-hot phase diffuse plasma of the Milky Way circumgalactic medium \citep{2023arXiv231010715L}. 
At fixed latitude $b$ within the background region, we compute the significance $S/N$ of the deabsorbed intensity\footnote{the deabsorbed \oo{} map is the observed \oo{} map multiplied by $e^\tau$, where $\tau=\sigma_{X}N_H$ is the optical depth produced by the X-ray absorbing column density $N_H$ through a cross-section $\sigma_X$. The absorbing layer is thus assumed to be in the foreground of the X-ray emitting plasma. We adopt the estimate for the total $N_H$ from \cite{2023arXiv231010715L}.} by first subtracting the estimated CGM emission at each latitude $b$
\begin{equation}
   \tilde{I} = I_X(l,b) - \langle I_{\rm cgm}(b) \rangle  
\end{equation}
where the notation ${}_{\rm cgm}(b)$ indicates a quantity sampled in the region limited by $220<l<250$ deg and $b\pm 5$ deg at a given $b$. The significance map $S/N$ is then obtained as
\begin{equation}
   \frac{S}{N} = \frac{ \tilde{I}(l,b) }{ {\rm std}( \tilde{I}_{\rm cgm}(b) ) }
\end{equation}

Fig.~\ref{fig:hmap_sigma} shows the derived $S/N$ map. 
While known bright and extended soft X-ray sources evidently pop up as coherent dark features in Fig.~\ref{fig:hmap_sigma} (e.g. SNRs, the Eridanus-Orion superbubble, the \erosita{} bubbles), around the LMC in pojection we found an extended patch of diffuse emission having yet no known counterpart.
We define the \myname{} as the coherent patch bounded by $S/N \geq 2$ around the LMC region. The region is approximately limited by the edge of the eROSITA bubble on the east side, while the northern bound extends all the way up to $b=-10$ deg, close to the Galactic disk (where the model of the absorption layer may introduce biases due to high column density values).

\subsection{X-ray line intensity ratio: temperature}
\begin{figure}
    \centering
    \includegraphics[width=0.49\textwidth]{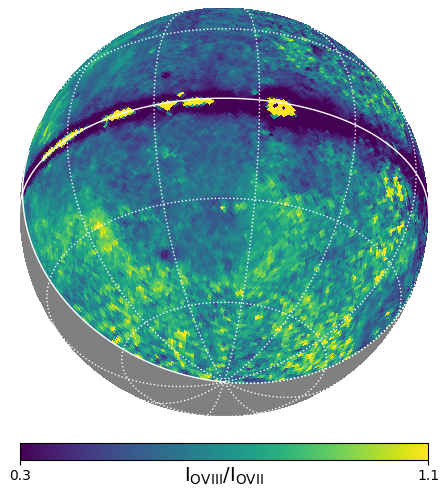}
    \caption{ \oo/\os{} temperature proxy map. 
    Contributions from the LHB foreground, the instrumental background, the CXB and the foreground absorption have been removed from the map before computing the ratio. See the text for further details. This map have been used to derive the map presented in Fig.~\ref{fig:temperature_maps}. }
    \label{fig:ratio_maps}
\end{figure}

The ratio of the \oo{} to \os{} CGM brightness has been converted into a proxy for the plasma temperature, and a first ever temperature map of the soft X-ray sky has been provided (Zheng et al., submitted). In Fig.~\ref{fig:ratio_maps}, we show the line ratio map obtained after corrections: background and foreground emission components have been subtracted from the narrowband maps of \os{} and \oo{}. The result was then deabsorbed, as described in \citet[][and Zheng et al. submitted]{2023arXiv231010715L} (we refer the reader here for details on the narrowband data preparation). We note that the conversion of the line ratio into a temperature associated to a plasma is biased in the case that the assumption of a single foreground absorption layer accounting for the total observed $N_H$ is broken. For instance, this possibly  happens in a circle of $\sim 2$ deg radius from the LMC and SMC center and across the Magellanic Bridge. In these regions a high HI column density $N_{HI}$ is observed and physically associated t the Magellanic Clouds, thus behind local and Galactic X-ray emission. These high $N_H$ regions within the Magellanic Clouds are currently not excluded from the computation of the foreground optical depth. Therefore, we note that the X-ray intensity and line ratio measured in the inner regions of the Magellanic Clouds and Bridge will be considered as biased and excluded from the analysis presented in this work. However, the excluded regions amount to only a few percent of the overall area covered by the \myname{} and will not affect the rest of our analysis, unless stated otherwise.

In addition, the \oo{} to \os{} line ratio can not be converted into a meaningful value of temperature whenever the two different line intensities involved in the ratio are produced by different sources or spectral components. 
For the reasons above and for completeness, we provide and refer the reader both to the line ratio value and to the temperature value corresponding to the ratio under the assumption of a single plamsa temperature (see Fig.~\ref{fig:temperature_maps}, referred to as "pseudo-temperature" $\mathcal{T}$ in Zheng et al. submitted), when potentially useful. We caution the reader that rather than an absolute calibration of the plasma temperature, the temperature map is mostly meant to show the approximate location and amplitude of potential temperature transitions, as the absolute normalization of the temperature relies on the details of the back-foreground subtraction operated on the narrowband images and is thus model-dependent.

\subsection{Optical reddening}

\cite{2019A&A...625A.135L} studied optical extinction in the Gaia data and produced a 3D map of the dust surrounding the Sun up to 600 pc in height and 3 kpc in distance along the plane parallel to the Milky Way stellar disk, centered on the position of the Sun. We converted the optical extinction into an equivalent hydrogen column density following the method proposed by \cite{Willingale.ea:13}. After converting the extinctions into $N_H$ in all the data cube, we re-project the cube onto \healpix{} grids \citep{2005ApJ...622..759G} at increasing distances from the Sun. At each distance step, the \healpix{} grid angular resolution is chosen to be the one closest to the angular size of a voxel in the original data cube by \citet{2019A&A...625A.135L} as seen from the Sun. Therefore, we can easily query the obtained Heapix maps iteratively in a given direction and reconstruct the $N_H$ differential (and cumulative) profile along the line of sight (LoS).

\subsection{Radio emission}

\begin{figure}
    \centering
    \includegraphics[width=0.49\textwidth]{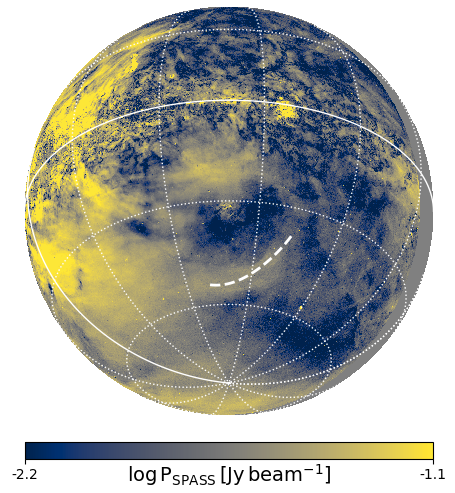}
    \caption{S-PASS 2.3 GHz radio polarization map \citep{2019MNRAS.489.2330C}. The white dashed line represents the location of the polarized radio loop XII as originally defined by \citet{2015MNRAS.452..656V}.}
    \label{fig:spass_P}
\end{figure}
\begin{figure}
    \centering
    \includegraphics[width=0.49\textwidth]{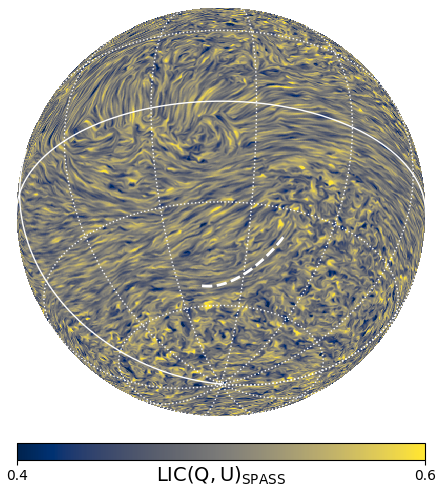}
    \caption{Linear Integral Convolution of the 23 GHz Planck Stokes parameter Q and U \citep{2016A&A...596A.109P}.}
    \label{fig:LIC_planck}
\end{figure}
In this work we use the S-Band Polarization All Sky Survey (S-PASS, \citealt{2019MNRAS.489.2330C}). This survey, run at 2.3 GHz by the Parkes telescope, has the advantage of combining a high sensitivity (0.81 mK beam$^{-1}$) to a wide coverage of the southern ecliptic hemisphere (Dec$<1$ deg, largely overlapping with the western Galactic hemisphere), therefore covering our field entirely. The S-PASS provides all Stokes product (I, Q, U and V). Additionally, the polarized radio emission (i.e. the linear combination of the Q and U Stokes $P = \sqrt{Q^2+U^2}$) is also provided. The S-PASS polarized intensity $P$ and polarization fraction $p=P/I$ are shown in Fig.~\ref{fig:spass_P} and Fig.~\ref{fig:spass_p}, respectively.

In addition, we consider the all-sky full-Stokes maps taken by the Planck satellite at 23 GHz \citep{2016A&A...596A.109P}. A Linear Integral Convolution (LIC) of the Q and U Stokes allows to show the direction of the magnetic field lines in terms of a color gradient (Fig.~\ref{fig:LIC_planck}). In principle, the LIC could also be performed on the S-PASS Stokes products. However, the S-PASS frequency of 2.3 GHz can be affected by Faraday rotation, an effect proportional to $\nu^{-2}$. The ten-fold higher frequency of Planck is instead negligibly affected by Galactic Faraday rotation and we consider it a more reliable proxy of the Galactic magnetic field direction. The LIC intensity gradients are perpendicular to the magnetic field direction. For example, the magnetic field is aligned along the extent of the radio loop XII, and is coincident with a higher polarization intensity and fraction in Fig.~\ref{fig:spass_P} and Fig.~\ref{fig:spass_p}, respectively.

\subsection{Absorption toward X-ray background sources}

Additional constraints on the physical properties of the hot plasma responsible for the X-ray emission in the \myname{} come from the study of high ionization lines (e.g. O VI, \os), when detected as absorption features in the UV and X-ray spectra of bright background sources. Previous studies on the black hole binary system LMC X-3 detected \os{} absorption line at redshift zero in that direction \citep{2005ApJ...635..386W, 2007ApJ...669..990B}. LMC X-3 is located 6.9 deg away (i.e. 6.7 kpc) from the center of the LMC.
PKS 0558-504 ($\rm (RA, DEC) = (05h\, 59m\, 47.38s,\, -50d\, 26m\, 52.4s),\, (l,b) = (257.96,\, -28.57)\, deg$), another source not far from the \myname{} but located in projection outside its boundaries, shows an \os{} absorption line of similar equivalent width to LMC X-3 \citep{2007ApJ...669..990B}.

\begin{figure}
    \centering
    \includegraphics[width=0.49\textwidth]{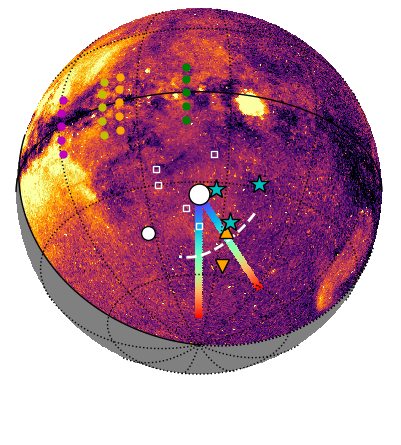}
    \caption{Chart map of the sources and features used and discussed in this work. The dotted lines perpendicular to the Galactic disk and centered at $b=0$ show directions tangential to the closest spiral arms of the Milky Way. By increasing $l$ (i.e. right to left) they represent the Carina (green dots, $l=284$ deg), Norma (orange dots, $l=306$ deg), Centaurus (yellow dots, $l=312$ deg) and Perseus (purple dots, $l=229$ deg) arms. The big white dots represent the SMC and LMC position. The empty white squares show the LoS characterized by high absorption shown in Fig.~\ref{fig:NH_vs_dist}. The cyan stars show the positions of the three bright X-ray sources with detected \os{} absorption line at $z=0$: PKS 0558-504, LMC X-3 and 1H 0419-577, listed by decreasing latitude. The dashed white line shows the position of radio loop XII. The rainbow lines show the data used to build the profiles and correlations in Fig.~\ref{fig:radial_profiles} and \ref{fig:correlations_all}. The orange triangles show the positions of the on-source (upward) and background (downward) diffuse emission spectra shown in Fig.~\ref{fig:spectrum}.}
    \label{fig:chart}
\end{figure}

With the aim of further testing the results obtained by \cite{2005ApJ...635..386W} and \cite{2007ApJ...669..990B} in additional and independent directions, we searched the XMM-Newton Science Archive\footnote{https://nxsa.esac.esa.int/nxsa-web/} for other bright soft X-ray sources found in projection within the boundaries of the \myname. We have found and analyzed all the observations with available data from the reflection grating spectrometer (RGS) onboard XMM-Newton, of the Seyfert 1 galaxy 1H 0419-577, collected between the years 2000-2018. The source is located at coordinates $\rm (RA,DEC)=(04h\, 26m\, 00.80s,\, -57d\, 12\arcmin\, 01.0\arcsec)\, or\, (l,b)=(266.987, -41.996)$ deg. The total available exposure covering the \os{} $z=0$ wavelength ($\rm \lambda_{OVII}=21.60\, \AA$) amounts to 369 ks. To the best of our knowledge, the X-ray spectrum of 1H 0419-577 has never been used to study Galactic \os{} absorption, as most of the observations have been carried after the thorough study of Galactic absorption by \cite{2007ApJ...669..990B}. By luck in this case, 1H 0419-577 also happens to be located in projection very close to the region where we have studied the X-ray spectrum of the excess emission, along the bright arc. The sky position of 1H 0419-577, LMC X-3 and PKS 0558-504 is shown in Fig.~\ref{fig:chart} (cyan stars in the lower right panel, respectively from the lowest latitude upward). We present the results of this analysis in the following section (Sec.~\ref{sec:results}, bullet point x).

\section{Results} \label{sec:results}

In this section we present different aspects of the observations, with small dependence on possible interpretations of their origin (these will be explored in Sec.~\ref{sec:discussion} instead). Given the number of different, yet important, findings, we list them into a series of bullet points in order to better distinguish between them and remind the reader of their complementary and independent flavor.

\begin{itemize}

\item[i)]{From the soft X-ray intensity maps (see Fig.~\ref{fig:hmap_O8cgm_n128}) we observe an arc-shaped brightening close to the position of the radio loop XII. The center of this arc is located in projection close to the LMC (cfr. with Fig.~\ref{fig:chart}), while the western end seems to point to the position of the SMC. These morphological features motivates us to investigate whether the \myname{} is physically connected to the LMC or not. We will discuss this hypothesis in Sec.~\ref{sec:discussion}. We also note the presence of an additional arc-shaped enhancement just southern than the brightest arc, tracing $\sim S/N\geq 2$. However, this second arc is much fainter. The bright X-ray arc seems to merge with the \myname{} edge (at $S/N\simeq 2$), on the eastern side. }

\begin{figure}
    \centering
    \includegraphics[width=0.49\textwidth]{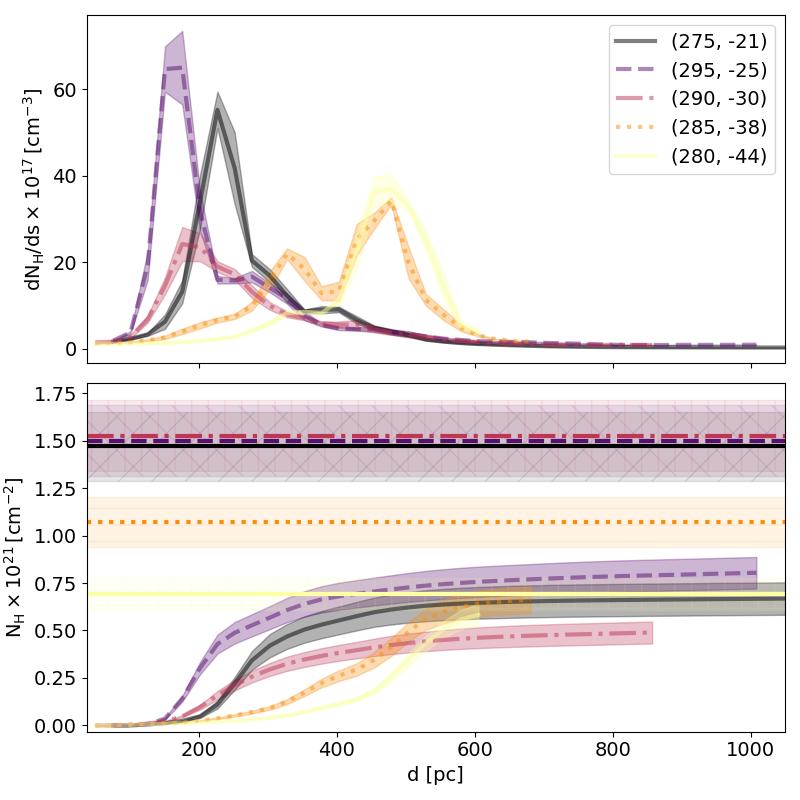}
    \caption{Column density $N_H$ of the X-ray absorbing material as a function of distance from the Sun, in the solar neighborhood. Top panel: differential $N_H$; Bottom panel: the solid lines show the cumulative $N_H$ for the same measurements presented in the top panel. The dashed lines and the hatched regions around them show the $N_H$ value with their uncertainty, as retrieved from the HI4PI survey in the same directions as indicated by the labelled colors in the top panel (see also the white empty squares in the chart map of Fig.~\ref{fig:chart}).}
    \label{fig:NH_vs_dist}
\end{figure}
\item[ii)]{Some absorbing clouds can be distinguished as a visual dimming of the X-ray intensity. These clouds can be used to put useful constraints (i.e. lower limits) to the distance of the \myname{}. From the dust extinction 3D data converted into $N_H$ estimates as explained above, we select LoS piercing some of the most evident absorbing clouds as shown in Fig.~\ref{fig:NH_vs_dist}. The peaks in the upper panel show the likely position of the X-ray absorbing clouds observed in front of the \myname{}. These clouds distance is as high as $\sim 500$ pc from the Sun. Therefore, we put a lower limit $d>500$ pc to the distance of the \myname{} from the Sun position. }

\begin{figure*}
    \centering
    \includegraphics[width=\textwidth]{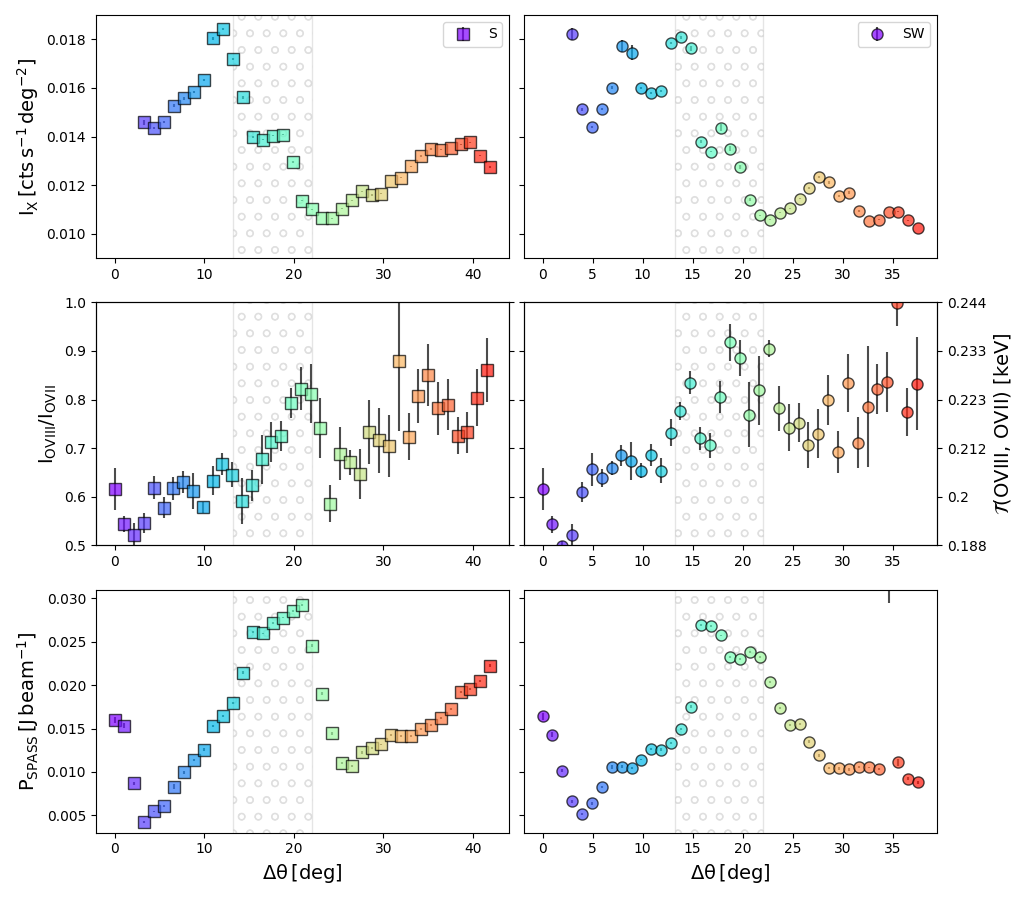}
    \caption{Radial profiles across the bright arc found within the \myname{}. The sky paths where the profiles have been extracted are shown in Fig.~\ref{fig:chart}. The rainbow colors match the data in this figure. $\Delta \theta \equiv 0$ is defined at coordinates $(l,b)=(280,-34)$. The hatched area shows the region dominated by a sharp intensity transition in the X-ray data. Top panel: \erosita{} soft X-ray intensity (0.5-1.0 keV); Central panel: temperature proxy (right vertical axis), as derived from the ratio of \erosita{} narrowband images centered around the \os{} and \oo{} emission lines (left vertical axis); 
    Bottom panel: polarization intensity $P$ along the same paths (S-PASS). The broad central peak matches the position of loop XII along the paths. }
    \label{fig:radial_profiles}
\end{figure*}
\item[iii)]{Driven by the morphology of the soft X-ray map, we investigate over the presence of a jump in the X-ray brightness across the bright arc. We extract a surface brightness profile from the paths shown in Fig.~\ref{fig:chart} in two radial directions chosen to be perpendicular to the arc curvature along the south (S) and south-west (SW) directions with respect to the arc centroid (i.e. close to the south ecliptic pole and LMC). The resulting profiles are shown in the top panels of Fig.~\ref{fig:radial_profiles}. The intensity values and their uncertainty are defined as the mean and the error on the mean inside each bin. A jump in the X-ray surface brightness at $\Delta\theta \sim 15$ deg shows the position of the bright arc in Fig.~\ref{fig:hmap_O8cgm_n128}. 
A second intensity jump (also briefly mentioned in bullet point i) seems to be present at $\Delta\theta\sim 18$ deg in both directions. 
We note that the detection of the brightness jumps does not depend on the choice of the intensity map to use (i.e. deabsorbed or not, right and left panels of Fig.~\ref{fig:hmap_O8cgm_n128}, respectively). This is thanks to the relatively low $N_H\leq 5\times 10^{20}$ value across the southern half of the \myname\ \citep[see][Fig. A.1]{2023arXiv231010715L}. Therefore, the intensity discontinuity is not generated by absorption.
}

\end{itemize}
\begin{figure}
    \centering
    \includegraphics[width=0.49\textwidth]{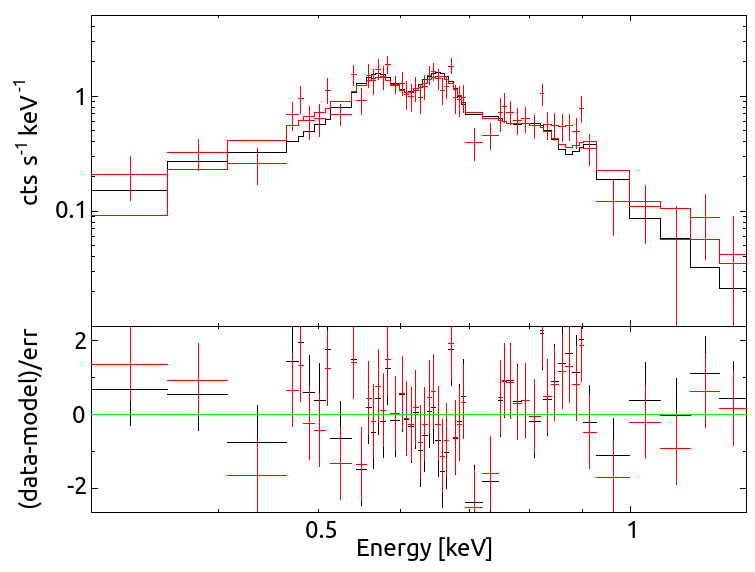}
    \caption{0.3-1.2 keV spectrum of the excess emission at the position of the bright arc. The spectrum from a background region from outside the \myname{} has been subtracted from the data. The on- and off-source (i.e. background) regions are indicated by the upward and downward orange triangles in Fig.~\ref{fig:chart}, respectively.}
    \label{fig:spectrum}
\end{figure}
\begin{figure*}
    \centering
    \includegraphics[width=\textwidth]{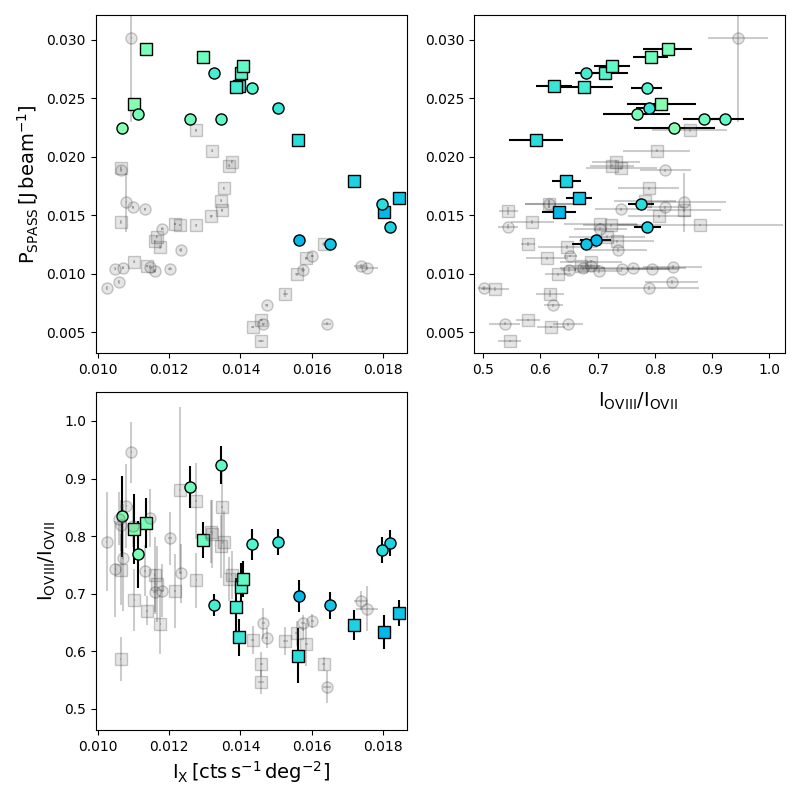}
    \caption{Scatter plots of the quantities presented in Fig.~\ref{fig:radial_profiles}. The symbols and colors of all the points correspond to the ones of Fig.~\ref{fig:radial_profiles}. Only the points included in the loop XII region (see the hatched region in Fig.~\ref{fig:radial_profiles}) are plotted with color in order to highlight their trends. }
    \label{fig:correlations_all}
\end{figure*}

Arc-shaped X-ray intensity jumps usually trace the position of shocks or cold fronts (hereafter, we refer to the set of all the mentioned classes as fronts). In order to distinguish between them, the temperature or entropy profiles is required. If the source of the X-ray emission is a hot thermal plasma with temperature $kT$ of about 0.1--1 keV. As previously discussed, a proxy of the temperature can be obtained by taking the ratio of \oo{} and \os{} line intensities. 

\begin{itemize}

\item[iv)]{We test the assumption of a thermal plasma by extracting soft band spectra from a region in coincidence of the bright arc emission (orange upward triangle in Fig.~\ref{fig:chart}, bottom right panel) and from outside the \myname{} (orange downward triangle). The latter is considered as a background reference and is subtracted from the former. We exploit the formal division of the \erosita\ data into sky squared "tiles" of 3 by 3 deg. The spectra from the \erosita\ sky tiles IDs 061147 and 044141 were extracted, respectively, and their difference was taken. 

The resulting spectrum in the soft X-ray range 0.3--1.2 keV, associated to the excess emission, is shown in Fig.~\ref{fig:spectrum}. The two bumps at around 0.57 and 0.65 keV immediately point toward a thermal nature of the source, as they likely trace \os{} and \oo{} line emission, to which \erosita{} has already been proved to be sensitive to \citep[][Zheng et al., submitted]{2023arXiv231010715L}. We fit the spectrum of the excess emission using a collisionally ionized plasma model
\footnote{\apec model, available with the \xspec{} software package and build on the AtomDB atomic database \citep{2001ApJ...556L..91S}.}.
We set $Z/Z_\odot$ to either 0.1 \citep{2023A&A...674A.195P} or 1.0. The black and red solid lines and residuals in Fig.~\ref{fig:spectrum} show two best-fits for different assumptions on the metal abundance of the hot plasma ($Z=1.0,\, 0.1\, Z_\odot$ respectively). The best-fits temperature is $0.21\pm 0.01$ and $0.19\pm 0.01$ keV respectively. 
The faint nature of the emission in the \myname, combined with the spectral energy resolution of \erosita, unfortunately prevents from constraining the metal abundance of the plasma. The fit does not strongly favor one of the assumed abundances (or temperatures). The metal abundance and the spectrum normalization are degenerate. The degeneracy is expected, as both parameters depend on the amount of underlying continuum emission, largely unconstrained by CCD instruments.

In addition, we tried a multi-temperature APEC model (GADEM). GADEM assumes a Gaussian distribution of temperatures. The best-fit temperature distribution peaks at the same temperature resulting from the simple APEC fit (with respective metal abundance) while holds the dispersion $\sigma_T=0.01$ keV of the temperature distribution is consistent with zero. The fit goodness does not improve with respect to APEC, despite the higher number of free parameters. We thus concluded that a single temperature model reasonably describes the excess emission, as observed by eROSITA, in the \myname{}. 

Regardless of the details of the fit of the excess emission, the evident \oo{} and \os{} features are enough to corroborate the hypothesis of a thermal plasma as the major source of the extended diffuse emission of the \myname{}.}

\item[v)]{From similar paths across the front, indicated by the rainbow paths in Fig.~\ref{fig:chart}, we built back-foreground corrected \oo/\os\ line ratio (Fig.~\ref{fig:ratio_maps}) as a proxY for temperature variations. We show the resulting profiles in the central panels of Fig.~\ref{fig:radial_profiles}.
Trends of increasing temperatures are observed at the location of the intensity jump (hatched region). The line ratio increase along the paths and across the hatched region amounts to $\sim 25\%$ and maps into a temperature increase of $\sim 10\%$. The line ratio (i.e. temperature) increase appears to be more significant in the south direction with respect to the south-west direction, due to a larger scatter of the data in the latter direction. However, a clear difference can be spotted once comparing the values before and after the hatched region in Fig.~\ref{fig:radial_profiles} encompassing the radio loop XII.
}

\item[vi)]{In general, the X-ray intensity can be interpreted as a proxy for $n^2 L$, where $n$ is the (average) density of the thermal plasma and $L$ is the length through the source along the LoS. 
By interpreting the line ratio as a proxy for the temperature, an anti-correlation between X-ray intensity (i.e. density) and the line ratio (i.e. temperature) then traces a contact discontinuity or cold front.
Cold fronts are surfaces of constant pressure where a density jump anti-correlates with a temperature difference (i.e. hotter on the low-density side of the front). On one hand, the anti-correlation is evident in the bottom left panel of Fig.~\ref{fig:correlations_all}. While we plot all the data points, we highlight the ones located across the bright X-ray jump and loop XII using the same colors and shapes as of Fig.~\ref{fig:radial_profiles}. The anti-correlation is also followed by the green to blue color gradient, indicating LoS piercing through the front by increasing distance to the origin. On the other hand, while the X-ray brightness in Fig.~\ref{fig:radial_profiles} shows a rather sharp transition, the temperature trend is milder and is harder to interpret it as a neat jump. Therefore, while the anti-correlation between intensity and line ratio is evident, it remains unclear whether the transition is tracing or not an actual cold front. The association with a cold front is therefore uncertain, yet possible.}

\item[vii)]{One of the previously known radio loops: radio loop XII, is now found to match the southern boundary of the \myname. To the best of our knowledge, this is the first time that this radio loop is connected to any other multiwavelength feature and studied. In particular, the northern edge of loop XII closely matches the bright X-ray arc across most of its length, while the southern edge of loop XII seems to match the $S/N\geq 2$ contour defining the \myname. A quantitative and complementary analysis of the polarized intensity (Fig.~\ref{fig:radial_profiles}, bottom panel) reveals the presence of a jump anti-correlating with the X-ray brightness across the same front presented above (Fig.~\ref{fig:correlations_all}, top left panel). The polarization jump is evident in both directions. Due to the anti-correlation between the X-ray intensity and line ratio, the anti-correlation between radio polarization and X-ray intensities may be recast in terms of a correlation between the radio polarization intensity and the temperature (Fig.~\ref{fig:correlations_all}), top right panel). The potential correlation however is not evident and dominated by a large scatter. 
Nevertheless, an enhanced radio polarized signal anti-correlating with the X-ray intensity jump also corroborates the presence of a cold front. }

\item[viii)]{In addition to the presence of enhanced polarized emission, we observe a magnetic direction following the length of the radio loop, therefore parallel to the front length, as already pointed out by \cite{2015MNRAS.452..656V}. The vectors direction is shown by a modulation of the intensity map given by the LIC of Stokes Q and U of the 23 GHz Planck data (see Fig.~\ref{fig:LIC_planck}). In principle we do not know the environment characterized by the detected magnetic field lines, while a Galactic origin is usually assumed given the coherent and very broad angular extent of the patterns. A Galactic origin of the polarized synchrotron emission is however assumed and corroborated in the studies of all the other known radio loops. The magnetic field direction in the region of the \myname{} other than radio loop XII is somewhat following the SE-NW direction, and is clearly displaced along the length of a dust absorbing cloud located in projection at about 10 deg north from the LMC. The cloud northern to the LMC can be easily spotted as the brightest central feature in the fractional polarization map $p=P/I$ of Fig.~\ref{fig:spass_p}, northern than the LMC. Outside the \myname the magnetic field morphology looks more disturbed and/or tangled compared to the regions within the boundaries of the \myname, strengthening the case for a coherent and distinct nature of the \myname.}

\item[ix)]{The polarized intensity in the S-PASS data shows a coherent decrease (to consistency with 0) over a broad circular patch in the central regions of the \myname{} (see the dark region in Fig.~\ref{fig:spass_P} and \ref{fig:spass_p}). This effect may be produced by physical depolarization of the signal, as the polarized intensity in this patch is even lower than the average level in the rest of the sky. We note that this patch is depolarized at 2.3 GHz but shows an ordered geometry of the magnetic field lines at 23 GHz.}

\end{itemize}

From the study of high ionization $z=0$ lines in the UV and X-ray spectra of the bright source LMC X-3 \citep{2007ApJ...669..990B}, happening to be in the background of the \myname, the observed wavelength of the \os{} absorption line at rest frame $21.592\pm 0.016 \AA$ was found to be consistent with a Galactic origin \citep{2007ApJ...669..990B}. Instead, this value is marginally inconsistent with the radial velocity of the LMC (expected at $21.623 \AA$), at the $2\sigma$ level. Similar results were previously obtained by \cite{2005ApJ...635..386W}. We note that the RGS nominal sensitivity in the \os{} band is $\rm 5\, m\AA$\footnote{https://heasarc.gsfc.nasa.gov/docs/xmm/uhb/rgs.html}, thus largely sufficient to probe a potential shift. The oxygen column density in the LMC X-3 direction was measured as $\rm EW_{OVII} = 21.0\pm 5.0\, m\AA$. In addition, the nonthermal broadening and the decreasing intensity (i.e. the scale height) of \os{} with respect to observed O VI suggest an origin of the high ionization states in a supernova-driven galactic fountain rather than in a hot atmosphere around the LMC \citep{2005ApJ...635..386W}. 
As already pointed out, PKS 0558-504, another source not far from the \myname{} but located in projection outside its boundaries, shows an \os{} absorption line of $\rm EW_{OVII}=21.7\pm 7.8\, m\AA$, similar to the one of LMC X-3.

\begin{itemize}

\item [x)]{In this work we add the constraints on the \os{} column density obtained in the direction of the bright source 1H 0419-577. By detecting the line and measuring an equivalent width of $\rm EW_{OVII}=42\pm 31\, m\AA$, we derive an oxygen column density of $\rm N_O=(8.4\pm 6.2)\times 10^{15}\, cm^{-2}$ (converted to $\rm N_H=(1.7\pm 1.3) \times 10^{19}\, cm^{-2}$ assuming a O/H ratio of $4.9\times 10^{-4}$, \citealt{2003ApJ...591.1220L}). We note that despite the separation of 12--16 deg between the lines of sight toward 1H 0419-577, LMC X-3 and PKS 0558-504, the values of the column density $\rm N_O$ are consistent with each other. However, as our measurement is still characterized by large uncertainties. A moderate amount of matter could still be hidden below our uncertainty values. }

\end{itemize}

\section{Discussion} \label{sec:discussion}

\subsection{A faint and old supernova remnant (alone) can not explain the X-ray intensity and extent}

The \myname{} is not unique in terms of extension and brightness when observing the X-ray sky. Other noticeable and similar regions are for instance the Eridanus-Orion superbubble \citep{2019A&A...631A..52J}, the Antlia and the Monogem Ring supernova remnants (SNRs, \citealt{Knies2022}). In addition, the position of the Antlia SNR for example shares a similar longitude range and (absolute) latitude with the \myname. Therefore, in this section we test the hypothesis that a single SNR can explain the emission of the \myname.

Above, we set a stringent lower limit to the distance of the hot plasma in the \myname, of $d> 500$ pc. A physical size of $\rm R\geq 131\, pc\, (\Theta/15 deg)(d/500 pc)$ is thus obtained. This value lays at about the upper bound of typical SNR sizes, but still is consistent with the possible range.
In the SNR scenario, we assume the bright arc to be the SNR shock surface, Thanks to its roughly circular shape. Despite having ruled out already a shock nature for the front, we apply a Sedov-Taylor formalism for shocks in SNRs as a test. The Sedov-Taylor equation links the density of the medium downstream the front with the age and the shock distance to the center of the SNR. From the observed X-ray excess emission in the \oo{} narrowband, given the radius $R$ as defined above and assuming a spherical geometry, we obtain an average density within the front of $\rm n\simeq 0.4-3.8\times 10^{-2}\, cm^{-3}$, where the quoted range includes the scatter on the observed surface brightness in the region downstream the front. By assuming $E=10^{51}$ erg as the total energy of the SNR, the age estimate derived by the Sedov-Taylor equation is $\Delta t=1.7-5.3\times 10^7$ yr. Now, the typical age up to which SNRs are X-ray bright is $\Delta t<10^5$ yr, thus incompatible with our estimate. 
A complementary approach is instead to compute the expected density using the upper limit to the age of $\Delta t=10^5$ yr. This keeps the timescale consistent with that of a faint and extended SNR located in the Milky Way halo (i.e. at high Galactic latitudes $|b|$). Under this assumption, we obtain $\rm n\sim 1\times 10^3\, cm^{-3}\, (R/131 pc)^5(\Delta t/10^5 yr)^{-2}$, that is unrealistically large for an old SNR. 

We stress that the results presented in this section are obtained for the closest possible distance to the SNR. In addition, $d=500$ pc for the SNR geometrical center is already at odds with the total observed $N_H$. The hypothesis that all the detected excess is produced by an old and faint SNR (located in the Milky Way halo) is thus unlikely. However, it is of course still possible that a single and very extended SNR may contribute to part of the observed emission and may be explored in future with multiwavelength data to look for other signatures typical of SNRs. The connection between X-ray and polarized radio intensity was in fact initially discovered by following this approach. So far we can only exclude that a single SNR can explain all of the properties of the \myname.

\subsection{A hot atmosphere around the Magellanic Clouds (alone) can not explain the X-ray intensity and extent}

As we noted above, the LMC is found (at least in projection) at the center of the \myname. The curvature of the bright X-ray arc and the loop XII also point toward a centroid close to (or overlapping with) the inner regions of the LMC itself. Cold, warm and warm-hot gas phases associated to the MCs have been detected as far as 45 deg from the LMC \citep{2009ApJS..181..398M, 2014ApJ...787..147F, 2017ApJ...851..110B, 2022Natur.609..915K}. These associations are usually based on the match of the radial velocity of emission lines with the one of the LMC ($v_{\rm LMC} 321\pm 24\, \rm km\, s^{-1}$, \citealt{2013ApJ...764..161K}), although the chances of a coincidence match are not negligible \citep{2015A&A...584L...6R}. The arguments above pose the question whether the \myname{} origin is connected to the Magellanic system (as a whole or in part).

On one hand, the detection of a warm-hot gas phase associated with the LMC and extended up to 45 deg from its center \citep{2020Natur.585..203L, 2022Natur.609..915K} sets the conditions for the presence of hotter gas in the case of a relatively broad distribution of temperatures \citep[e.g.,][]{2021AJ....161...57G}. In addition, discontinuities in the gas properties produced by the interaction of the Magellanic Clouds with the circumgalactic medium (CGM) of the Milky Way, as well as the mutual interaction between the Small and Large Magellanic Clouds, are plausible \citep{2010ApJ...721L..97B, 2023arXiv230810963S} and observed \citep[e.g.,][]{2016ARA&A..54..363D, 2021Natur.592..534C}. In this scenario, the front detected in the \myname{} could then be caused at the interface between the Magellanic hot corona and the CGM of the Milky Way due to the three body interaction. Although its position at the opposite side of the direction of the LMC motion with respect to the Milky Way, a correlation is not expected in general \citep{2006ApJ...650..102A} and a different front may still be located in the leading direction while remaining hidden behind the thick dust clouds observed in the north of the \myname. We also note that shocks and cold fronts are known to compress and align magnetic field lines, thus providing a possible explanation for the loop XII radio emission and its anti-correlation with the X-ray intensity. Although this is interesting however, we note that the presence of the radio emission does not allow to discriminate between this particular scenario and a front with similar observed properties but produced by a smaller-scale (i.e. Galactic) phenomenon.

On the other hand, as mentioned above, studies of the high ionization lines (O VI, \os, \oo{} and Ne IX) detected in absorption are inconsistent with the LMC properties in terms of redshift and dispersion \citep{2005ApJ...635..386W, 2007ApJ...669..990B}. A new line of sight presented in this work, toward the quasar 1H 0419-577 shows an \os{} absorption line with an equivalent width of about $\rm EW=42\pm 31\, m\AA$, consistent with the values observed toward LMC X-3 and PKS 0558-504. The lines of sight toward these different object are separated by 12--16 deg. The similarity of the measured equivalent widths (and in turn of the column density) can be explained if the major contribution to the $z=0$ \os{} absorption line in the spectrum of the background sources comes from hot gas of Galactic origin, spread over a broad region. This suggests that the bulk of the observed X-ray emission of the \myname{} is associated with a Galactic origin. We note however that a small contribution ($\rm N_O<10^{15}\, cm^{-2}$) from a hot Magellanic corona may be still possible.

Overall, hot gas in the Magellanic system, although at odds with providing a viable explanation for the \myname{} as a whole, may be still connected to some of its features. For instance, it could account for part of the observed emission or be connected to the front. Unfortunately, concerning the soft X-ray emission detected in this work and characterized by prominent \os{} and \oo{} lines, the energy resolution of \erosita{} prevents from detecting a line shift or broadening of the diffuse emission. Recent and future instruments probing soft X-ray line width with a $\sim$eV resolution (e.g., XRISM, Athena, LEM) will be crucial to further test this scenario. 

\subsection{Possible connection to star-formation in the disk}

In regions of similar brightness and/or extent (i.e. large angular scales), the faint X-ray emission has been explained by the overlap of several SNRs and/or winds from hot stars. One notable example apart from single SNRs is the Eridanus-Orion superbubble. Superbubbles are connected to SNRs and hot OB star associations usually found in giant molecular clouds \citep{2015ApJS..216...18N, 2020A&A...639A.110A} and have already been associated to radio loops \citep{2023A&A...677L..11B}. A superbubble is thus a reasonable candidate as an explanation of the \myname. Therefore, we checked for the location of HII star forming nearby regions. HII regions are usually found in the densest parts of the Galactic disk (i.e. at low Galactic absolute latitudes). A recent survey of these regions provides information on their distribution along the Galactic plane \citep{2014ApJS..212....1A}. HII regions are not detected nearby the center of the \myname{} (i.e. nearby the LMC). However, they may be located at smaller latitudes in projection. A sharp decrease at around l=300-330 deg along Galactic longitudes is observed in the distribution of the HII regions.
A recent catalog of OB associations found these structures distributed all along the Galactic disk in projection. In particular, the longitude range 280--290 exhibits a small concentration of OB associations, thus potentially connected to the X-ray emission in the \myname{}.
In general by assuming a spherical geometry, the \myname{} physical diameter is equal to its distance from us given its apparent angular size $\Delta\theta \sim 45$ deg. The OB star associations at similar longitude are located in a range of 2--3 kpc from the Sun. This would imply a physical radius of 1--1.5 kpc. We note that this size is at odds (i.e. much larger) than the size of typical superbubbles of $\sim 10^2$ pc.
Another tracer for warm ionized gas produced in superbubbles is the $H\alpha$ line emission. The Eri-Ori superbubble is clearly detected in large angular scale maps of this tracer \citep{2001PASP..113.1326G}. In contrast, neither the \erosita{} bubbles nor the \myname{} are detected in the same map. However, this may be explained by the larger distance (i.e. fainter emission) of the \myname{} from us compared to the Eri-Ori superbubble.

Other than superbubbles, regions of enhanced star formation activity in the disk of the Milky Way have been mapped through a cross match between massive young stellar objects, HII regions and methanol masers \citep{2014MNRAS.443.1555U}, producing a map of the clumps of massive stars in the disk of the Milky Way. Clumps of massive star are found at Galactic longitudes as low as 280 deg, similarly to the OB associations.
Unfortunately, the eastern bound of the \myname{} is unconstrained due to the presence of brighter emission produced by the western edge of the southern \erosita{} bubble at around $l=290$ deg. Therefore, we can only put an upper limit to the Galactic longitude reached by the \myname{} of $l<290$ deg. This limit is close to the edge of the longitude range of active star forming region, but still consistent with it.

An additional possibility lies in the connection with gas heated and expelled by the stellar formation and evolution activity into the spiral arms of the Milky Way. A recent map of the closest arms \citep{2023ApJ...947...54X} provides the direction where the LoS is tangential to the arms. We show their positions in the chart map of Fig.~\ref{fig:chart}. The Carina spiral arm seems to match the longitude range of the Goat Horn. If related to it, the relative distance of 3-4 kpc to this arm would also imply a similar size of the Goat Horn complex, thus reaching heights smaller, yet comparable, to the ones of the \erosita{} bubble. In this scenario, the apparent (i.e. projected) connection of radio loop XII with the central regions of the Milky Way could be explained by the actual proximity of these regions.

Despite the connections with star forming regions in the disk of the Milky Way, a strong physical link with the emission of the \myname{} can not be confirmed nor ruled out at the moment.

\section{Conclusion} \label{sec:conclusion}

Below, we summarize what are the new observations and results discussed in this work:
\begin{itemize}

\item a large and coherent patch of soft X-ray emission exceeding the background, extending up to $\sim 1000$ deg$^2$ is found around the Dorado region. This feature (dubbed as the \myname{}) has not been previously recognized as a coherent and separate structure of the sky background;
\item we observe foreground dust clouds absorbing the soft X-ray emission. These clouds are found as far as 500 pc from the Sun, given the available 3D extinction data;
\item an additional brightening of the X-ray emission, characterized by an arc-shaped morphology centered on about the LMC position is blended with the emission from the \myname{}, at least in projection;
\item we find two X-ray intensity jumps across the boundary of the enhanced X-ray feature (i.e. a front);
\item the spectrum of the excess emission at the front is consistent with the one produced by a collisionally ionized hot plasma;
\item the region delimiting the front (north) and the boundary of the \myname{} (south) coincide with an arc-shaped radio polarization enhancement previously known as radio loop XII;
\item we find a temperature gradient increase across the front, anti-correlating with the X-ray intensity. 
\item an anti-correlation between the X-ray brightness and the polarized radio intensity is also detected across the loop XII. Both the detected anti-correlations are characteristic of cold fronts;
\item the lower limit to the distance prevents the \myname{} to be associated to a single supernova remnant;
\item the morphology of the Galactic magnetic field lines is aligned with the radio loop XII (i.e. upstream the front) and along a foreground dust cloud. The lines are less ordered in the region outside the boundaries of the \myname{}. The center of the \myname{} is characterized by a very low fractional polarization, which can be due to physical (Faraday) depolarization. Depolarization and tangled magnetic field lines usually characterise turbulent magnetic fields;
\item absorption of X-ray photons by bright quasar due to $O_6^{+}$ ions is small. These measurements translate to tight upper limits to the \os{} column density, leaving little room for the presence of an extended hot atmosphere around the LMC in order to explain the observed X-ray emission from the \myname{};

\end{itemize}

Our work discussed different possible interpretations of the \myname.
It is still possible for the emission of the \myname{} to be explained by a combinations of the models presented above. Future measurements of the width of the high ionization lines emitting X-rays thanks to upcoming missions (e.g. XRISM, LEM, Athena) will be crucial in order to disentangle between different emission components and better constrain the source distance.

{\small \paragraph{Acknowledgements}
This work is based on data from eROSITA, the soft X-ray instrument aboard SRG, a joint Russian-German science mission supported by the Russian Space Agency (Roskosmos), in the interests of the Russian Academy of Sciences represented by its Space Research Institute (IKI), and the Deutsches Zentrum für Luft- und Raumfahrt (DLR). The SRG spacecraft was built by Lavochkin Association (NPOL) and its subcontractors, and is operated by NPOL with support from the Max Planck Institute for Extraterrestrial Physics (MPE). The development and construction of the eROSITA X-ray instrument was led by MPE, with contributions from the Dr. Karl Remeis Observatory Bamberg \& ECAP (FAU Erlangen-Nuernberg), the University of Hamburg Observatory, the Leibniz Institute for Astrophysics Potsdam (AIP), and the Institute for Astronomy and Astrophysics of the University of Tübingen, with the support of DLR and the Max Planck Society. The Argelander Institute for Astronomy of the University of Bonn and the Ludwig Maximilians Universität Munich also participated in the science preparation for eROSITA.
The eROSITA data shown here were processed using the eSASS software system developed by the German eROSITA consortium.
NL, GP and XZ acknowledge financial support from the European Research Council (ERC) under the European Union’s Horizon 2020 research and innovation program HotMilk (grant agreement No. [865637]).
GP also aknowledges support from Bando per il Finanziamento della Ricerca Fondamentale 2022 dell’Istituto Nazionale di Astrofisica (INAF): GO Large program and from the Framework per l’Attrazione e il Rafforzamento delle Eccellenze (FARE) per la ricerca in Italia (R20L5S39T9).}

\bibliographystyle{aa}
\bibliography{aa} 

\begin{thebibliography}{44}
\expandafter\ifx\csname natexlab\endcsname\relax\def\natexlab#1{#1}\fi

\bibitem[{{Abdullah} \& {Tielens}(2020)}]{2020A&A...639A.110A}
{Abdullah}, A. \& {Tielens}, A.~G.~G.~M. 2020, \aap, 639, A110

\bibitem[{{Anderson} {et~al.}(2014){Anderson}, {Bania}, {Balser}, {Cunningham}, {Wenger}, {Johnstone}, \& {Armentrout}}]{2014ApJS..212....1A}
{Anderson}, L.~D., {Bania}, T.~M., {Balser}, D.~S., {et~al.} 2014, \apjs, 212, 1

\bibitem[{{Ascasibar} \& {Markevitch}(2006)}]{2006ApJ...650..102A}
{Ascasibar}, Y. \& {Markevitch}, M. 2006, \apj, 650, 102

\bibitem[{{Barger} {et~al.}(2017){Barger}, {Madsen}, {Fox}, {Wakker}, {Bland-Hawthorn}, {Nidever}, {Haffner}, {Antwi-Danso}, {Hernandez}, {Lehner}, {Hill}, {Curzons}, \& {Tepper-Garc{\'\i}a}}]{2017ApJ...851..110B}
{Barger}, K.~A., {Madsen}, G.~J., {Fox}, A.~J., {et~al.} 2017, \apj, 851, 110

\bibitem[{{Besla} {et~al.}(2010){Besla}, {Kallivayalil}, {Hernquist}, {van der Marel}, {Cox}, \& {Kere{\v{s}}}}]{2010ApJ...721L..97B}
{Besla}, G., {Kallivayalil}, N., {Hernquist}, L., {et~al.} 2010, \apjl, 721, L97

\bibitem[{{Bracco} {et~al.}(2023){Bracco}, {Padovani}, \& {Soler}}]{2023A&A...677L..11B}
{Bracco}, A., {Padovani}, M., \& {Soler}, J.~D. 2023, \aap, 677, L11

\bibitem[{{Bregman} \& {Lloyd-Davies}(2007)}]{2007ApJ...669..990B}
{Bregman}, J.~N. \& {Lloyd-Davies}, E.~J. 2007, \apj, 669, 990

\bibitem[{{Carretti} {et~al.}(2019){Carretti}, {Haverkorn}, {Staveley-Smith}, {Bernardi}, {Gaensler}, {Kesteven}, {Poppi}, {Brown}, {Crocker}, {Purcell}, {Schnitzeler}, \& {Sun}}]{2019MNRAS.489.2330C}
{Carretti}, E., {Haverkorn}, M., {Staveley-Smith}, L., {et~al.} 2019, \mnras, 489, 2330

\bibitem[{{Conroy} {et~al.}(2021){Conroy}, {Naidu}, {Garavito-Camargo}, {Besla}, {Zaritsky}, {Bonaca}, \& {Johnson}}]{2021Natur.592..534C}
{Conroy}, C., {Naidu}, R.~P., {Garavito-Camargo}, N., {et~al.} 2021, \nat, 592, 534

\bibitem[{{D'Onghia} \& {Fox}(2016)}]{2016ARA&A..54..363D}
{D'Onghia}, E. \& {Fox}, A.~J. 2016, \araa, 54, 363

\bibitem[{{Fox} {et~al.}(2014){Fox}, {Wakker}, {Barger}, {Hernandez}, {Richter}, {Lehner}, {Bland-Hawthorn}, {Charlton}, {Westmeier}, {Thom}, {Tumlinson}, {Misawa}, {Howk}, {Haffner}, {Ely}, {Rodriguez-Hidalgo}, \& {Kumari}}]{2014ApJ...787..147F}
{Fox}, A.~J., {Wakker}, B.~P., {Barger}, K.~A., {et~al.} 2014, \apj, 787, 147

\bibitem[{{Freyberg}(1998)}]{1998LNP...506..113F}
{Freyberg}, M.~J. 1998, in IAU Colloq. 166: The Local Bubble and Beyond, ed. D.~{Breitschwerdt}, M.~J. {Freyberg}, \& J.~{Truemper}, Vol. 506, 113--116

\bibitem[{{Gaustad} {et~al.}(2001){Gaustad}, {McCullough}, {Rosing}, \& {Van Buren}}]{2001PASP..113.1326G}
{Gaustad}, J.~E., {McCullough}, P.~R., {Rosing}, W., \& {Van Buren}, D. 2001, \pasp, 113, 1326

\bibitem[{{G{\'o}rski} {et~al.}(2005){G{\'o}rski}, {Hivon}, {Banday}, {Wandelt}, {Hansen}, {Reinecke}, \& {Bartelmann}}]{2005ApJ...622..759G}
{G{\'o}rski}, K.~M., {Hivon}, E., {Banday}, A.~J., {et~al.} 2005, \apj, 622, 759

\bibitem[{{Gulick} {et~al.}(2021){Gulick}, {Kaaret}, {Zajczyk}, {LaRocca}, {Bluem}, {Ringuette}, {Jahoda}, \& {Kuntz}}]{2021AJ....161...57G}
{Gulick}, H., {Kaaret}, P., {Zajczyk}, A., {et~al.} 2021, \aj, 161, 57

\bibitem[{{Joubaud} {et~al.}(2019){Joubaud}, {Grenier}, {Ballet}, \& {Soler}}]{2019A&A...631A..52J}
{Joubaud}, T., {Grenier}, I.~A., {Ballet}, J., \& {Soler}, J.~D. 2019, \aap, 631, A52

\bibitem[{{Kallivayalil} {et~al.}(2013){Kallivayalil}, {van der Marel}, {Besla}, {Anderson}, \& {Alcock}}]{2013ApJ...764..161K}
{Kallivayalil}, N., {van der Marel}, R.~P., {Besla}, G., {Anderson}, J., \& {Alcock}, C. 2013, \apj, 764, 161

\bibitem[{Knies(2022)}]{Knies2022}
Knies, J.~R. 2022, doctoralthesis, Friedrich-Alexander-Universit{\"a}t Erlangen-N{\"u}rnberg (FAU)

\bibitem[{{Krishnarao} {et~al.}(2022){Krishnarao}, {Fox}, {D'Onghia}, {Wakker}, {Cashman}, {Howk}, {Lucchini}, {French}, \& {Lehner}}]{2022Natur.609..915K}
{Krishnarao}, D., {Fox}, A.~J., {D'Onghia}, E., {et~al.} 2022, \nat, 609, 915

\bibitem[{{Lallement} {et~al.}(2019){Lallement}, {Babusiaux}, {Vergely}, {Katz}, {Arenou}, {Valette}, {Hottier}, \& {Capitanio}}]{2019A&A...625A.135L}
{Lallement}, R., {Babusiaux}, C., {Vergely}, J.~L., {et~al.} 2019, \aap, 625, A135

\bibitem[{{Locatelli} {et~al.}(2024){Locatelli}, {Ponti}, {Zheng}, {Merloni}, {Becker}, {Comparat}, {Dennerl}, {Freyberg}, {Sasaki}, \& {Yeung}}]{2023arXiv231010715L}
{Locatelli}, N., {Ponti}, G., {Zheng}, X., {et~al.} 2024, \aap, 681, A78

\bibitem[{{Lodders}(2003)}]{2003ApJ...591.1220L}
{Lodders}, K. 2003, \apj, 591, 1220

\bibitem[{{Lucchini} {et~al.}(2020){Lucchini}, {D'Onghia}, {Fox}, {Bustard}, {Bland-Hawthorn}, \& {Zweibel}}]{2020Natur.585..203L}
{Lucchini}, S., {D'Onghia}, E., {Fox}, A.~J., {et~al.} 2020, \nat, 585, 203

\bibitem[{{McClure-Griffiths} {et~al.}(2009){McClure-Griffiths}, {Pisano}, {Calabretta}, {Ford}, {Lockman}, {Staveley-Smith}, {Kalberla}, {Bailin}, {Dedes}, {Janowiecki}, {Gibson}, {Murphy}, {Nakanishi}, \& {Newton-McGee}}]{2009ApJS..181..398M}
{McClure-Griffiths}, N.~M., {Pisano}, D.~J., {Calabretta}, M.~R., {et~al.} 2009, \apjs, 181, 398

\bibitem[{{Merloni} {et~al.}(2024){Merloni}, {Lamer}, {Liu}, {Ramos-Ceja}, {Brunner}, {Bulbul}, {Dennerl}, \& et~al.}]{Merloni24}
{Merloni}, A., {Lamer}, G., {Liu}, T., {et~al.} 2024, \aap, 682, A34

\bibitem[{{Nishimura} {et~al.}(2015){Nishimura}, {Tokuda}, {Kimura}, {Muraoka}, {Maezawa}, {Ogawa}, {Dobashi}, {Shimoikura}, {Mizuno}, {Fukui}, \& {Onishi}}]{2015ApJS..216...18N}
{Nishimura}, A., {Tokuda}, K., {Kimura}, K., {et~al.} 2015, \apjs, 216, 18

\bibitem[{{Planck Collaboration} {et~al.}(2016){Planck Collaboration}, {Aghanim}, {Ashdown}, {Aumont}, {Baccigalupi}, {Ballardini}, {Banday}, {Barreiro}, {Bartolo}, {Basak}, {Benabed}, {Bernard}, {Bersanelli}, {Bielewicz}, {Bonavera}, {Bond}, {Borrill}, {Bouchet}, {Boulanger}, {Burigana}, {Calabrese}, {Cardoso}, {Carron}, {Chiang}, {Colombo}, {Comis}, {Couchot}, {Coulais}, {Crill}, {Curto}, {Cuttaia}, {de Bernardis}, {de Zotti}, {Delabrouille}, {Di Valentino}, {Dickinson}, {Diego}, {Dor{\'e}}, {Douspis}, {Ducout}, {Dupac}, {Dusini}, {Elsner}, {En{\ss}lin}, {Eriksen}, {Falgarone}, {Fantaye}, {Finelli}, {Forastieri}, {Frailis}, {Fraisse}, {Franceschi}, {Frolov}, {Galeotta}, {Galli}, {Ganga}, {G{\'e}nova-Santos}, {Gerbino}, {Ghosh}, {Giraud-H{\'e}raud}, {Gonz{\'a}lez-Nuevo}, {G{\'o}rski}, {Gruppuso}, {Gudmundsson}, {Hansen}, {Helou}, {Henrot-Versill{\'e}}, {Herranz}, {Hivon}, {Huang}, {Jaffe}, {Jones}, {Keih{\"a}nen}, {Keskitalo}, {Kiiveri}, {Kisner}, {Krachmalnicoff}, {Kunz}, {Kurki-Suonio}, {Lamarre},
  {Langer}, {Lasenby}, {Lattanzi}, {Lawrence}, {Le Jeune}, {Levrier}, {Lilje}, {Lilley}, {Lindholm}, {L{\'o}pez-Caniego}, {Ma}, {Mac{\'\i}as-P{\'e}rez}, {Maggio}, {Maino}, {Mandolesi}, {Mangilli}, {Maris}, {Martin}, {Mart{\'\i}nez-Gonz{\'a}lez}, {Matarrese}, {Mauri}, {McEwen}, {Melchiorri}, {Mennella}, {Migliaccio}, {Miville-Desch{\^e}nes}, {Molinari}, {Moneti}, {Montier}, {Morgante}, {Moss}, {Natoli}, {Oxborrow}, {Pagano}, {Paoletti}, {Patanchon}, {Perdereau}, {Perotto}, {Pettorino}, {Piacentini}, {Plaszczynski}, {Polastri}, {Polenta}, {Puget}, {Rachen}, {Racine}, {Reinecke}, {Remazeilles}, {Renzi}, {Rocha}, {Rosset}, {Rossetti}, {Roudier}, {Rubi{\~n}o-Mart{\'\i}n}, {Ruiz-Granados}, {Salvati}, {Sandri}, {Savelainen}, {Scott}, {Sirignano}, {Sirri}, {Soler}, {Spencer}, {Suur-Uski}, {Tauber}, {Tavagnacco}, {Tenti}, {Toffolatti}, {Tomasi}, {Tristram}, {Trombetti}, {Valiviita}, {Van Tent}, {Vielva}, {Villa}, {Vittorio}, {Wandelt}, {Wehus}, {Zacchei}, \& {Zonca}}]{2016A&A...596A.109P}
{Planck Collaboration}, {Aghanim}, N., {Ashdown}, M., {et~al.} 2016, \aap, 596, A109

\bibitem[{{Ponti} {et~al.}(2019){Ponti}, {Hofmann}, {Churazov}, {Morris}, {Haberl}, {Nandra}, {Terrier}, {Clavel}, \& {Goldwurm}}]{2019Natur.567..347P}
{Ponti}, G., {Hofmann}, F., {Churazov}, E., {et~al.} 2019, \nat, 567, 347

\bibitem[{{Ponti} {et~al.}(2021){Ponti}, {Morris}, {Churazov}, {Heywood}, \& {Fender}}]{2021A&A...646A..66P}
{Ponti}, G., {Morris}, M.~R., {Churazov}, E., {Heywood}, I., \& {Fender}, R.~P. 2021, \aap, 646, A66

\bibitem[{{Ponti} {et~al.}(2023){Ponti}, {Zheng}, {Locatelli}, {Bianchi}, {Zhang}, {Anastasopoulou}, {Comparat}, {Dennerl}, {Freyberg}, {Haberl}, {Merloni}, {Reiprich}, {Salvato}, {Sanders}, {Sasaki}, {Strong}, \& {Yeung}}]{2023A&A...674A.195P}
{Ponti}, G., {Zheng}, X., {Locatelli}, N., {et~al.} 2023, \aap, 674, A195

\bibitem[{{Predehl} {et~al.}(2021){Predehl}, {Andritschke}, {Arefiev}, {Babyshkin}, {Batanov}, {Becker}, {B{\"o}hringer}, {Bogomolov}, {Boller}, {Borm}, {Bornemann}, {Br{\"a}uninger}, {Br{\"u}ggen}, {Brunner}, {Brusa}, {Bulbul}, {Buntov}, {Burwitz}, {Burkert}, {Clerc}, {Churazov}, {Coutinho}, {Dauser}, {Dennerl}, {Doroshenko}, {Eder}, {Emberger}, {Eraerds}, {Finoguenov}, {Freyberg}, {Friedrich}, {Friedrich}, {F{\"u}rmetz}, {Georgakakis}, {Gilfanov}, {Granato}, {Grossberger}, {Gueguen}, {Gureev}, {Haberl}, {H{\"a}lker}, {Hartner}, {Hasinger}, {Huber}, {Ji}, {Kienlin}, {Kink}, {Korotkov}, {Kreykenbohm}, {Lamer}, {Lomakin}, {Lapshov}, {Liu}, {Maitra}, {Meidinger}, {Menz}, {Merloni}, {Mernik}, {Mican}, {Mohr}, {M{\"u}ller}, {Nandra}, {Nazarov}, {Pacaud}, {Pavlinsky}, {Perinati}, {Pfeffermann}, {Pietschner}, {Ramos-Ceja}, {Rau}, {Reiffers}, {Reiprich}, {Robrade}, {Salvato}, {Sanders}, {Santangelo}, {Sasaki}, {Scheuerle}, {Schmid}, {Schmitt}, {Schwope}, {Shirshakov}, {Steinmetz}, {Stewart}, {Str{\"u}der},
  {Sunyaev}, {Tenzer}, {Tiedemann}, {Tr{\"u}mper}, {Voron}, {Weber}, {Wilms}, \& {Yaroshenko}}]{2021A&A...647A...1P}
{Predehl}, P., {Andritschke}, R., {Arefiev}, V., {et~al.} 2021, \aap, 647, A1

\bibitem[{{Predehl} {et~al.}(2020){Predehl}, {Sunyaev}, {Becker}, {Brunner}, {Burenin}, {Bykov}, {Cherepashchuk}, {Chugai}, {Churazov}, {Doroshenko}, {Eismont}, {Freyberg}, {Gilfanov}, {Haberl}, {Khabibullin}, {Krivonos}, {Maitra}, {Medvedev}, {Merloni}, {Nandra}, {Nazarov}, {Pavlinsky}, {Ponti}, {Sanders}, {Sasaki}, {Sazonov}, {Strong}, \& {Wilms}}]{2020Natur.588..227P}
{Predehl}, P., {Sunyaev}, R.~A., {Becker}, W., {et~al.} 2020, \nat, 588, 227

\bibitem[{{Richter} {et~al.}(2015){Richter}, {de Boer}, {Werner}, \& {Rauch}}]{2015A&A...584L...6R}
{Richter}, P., {de Boer}, K.~S., {Werner}, K., \& {Rauch}, T. 2015, \aap, 584, L6

\bibitem[{{Robertson} \& {Cravens}(2003)}]{2003JGRA..108.8031R}
{Robertson}, I.~P. \& {Cravens}, T.~E. 2003, Journal of Geophysical Research (Space Physics), 108, 8031

\bibitem[{{Setton} {et~al.}(2023){Setton}, {Besla}, {Patel}, {Hummels}, {Zheng}, \& {Schneider}}]{2023arXiv230810963S}
{Setton}, D.~J., {Besla}, G., {Patel}, E., {et~al.} 2023, arXiv e-prints, arXiv:2308.10963

\bibitem[{{Smith} {et~al.}(2001){Smith}, {Brickhouse}, {Liedahl}, \& {Raymond}}]{2001ApJ...556L..91S}
{Smith}, R.~K., {Brickhouse}, N.~S., {Liedahl}, D.~A., \& {Raymond}, J.~C. 2001, \apjl, 556, L91

\bibitem[{{Snowden} \& {Schmitt}(1990)}]{1990Ap&SS.171..207S}
{Snowden}, S.~L. \& {Schmitt}, J.~H.~M.~M. 1990, \apss, 171, 207

\bibitem[{{Sunyaev} {et~al.}(2021){Sunyaev}, {Arefiev}, {Babyshkin}, {Bogomolov}, {Borisov}, {Buntov}, {Brunner}, {Burenin}, {Churazov}, {Coutinho}, {Eder}, {Eismont}, {Freyberg}, {Gilfanov}, {Gureyev}, {Hasinger}, {Khabibullin}, {Kolmykov}, {Komovkin}, {Krivonos}, {Lapshov}, {Levin}, {Lomakin}, {Lutovinov}, {Medvedev}, {Merloni}, {Mernik}, {Mikhailov}, {Molodtsov}, {Mzhelsky}, {M{\"u}ller}, {Nandra}, {Nazarov}, {Pavlinsky}, {Poghodin}, {Predehl}, {Robrade}, {Sazonov}, {Scheuerle}, {Shirshakov}, {Tkachenko}, \& {Voron}}]{2021A&A...656A.132S}
{Sunyaev}, R., {Arefiev}, V., {Babyshkin}, V., {et~al.} 2021, \aap, 656, A132

\bibitem[{{Urquhart} {et~al.}(2014){Urquhart}, {Moore}, {Csengeri}, {Wyrowski}, {Schuller}, {Hoare}, {Lumsden}, {Mottram}, {Thompson}, {Menten}, {Walmsley}, {Bronfman}, {Pfalzner}, {K{\"o}nig}, \& {Wienen}}]{2014MNRAS.443.1555U}
{Urquhart}, J.~S., {Moore}, T.~J.~T., {Csengeri}, T., {et~al.} 2014, \mnras, 443, 1555

\bibitem[{{Vidal} {et~al.}(2015){Vidal}, {Dickinson}, {Davies}, \& {Leahy}}]{2015MNRAS.452..656V}
{Vidal}, M., {Dickinson}, C., {Davies}, R.~D., \& {Leahy}, J.~P. 2015, \mnras, 452, 656

\bibitem[{{Wang} {et~al.}(2005){Wang}, {Yao}, {Tripp}, {Fang}, {Cui}, {Nicastro}, {Mathur}, {Williams}, {Song}, \& {Croft}}]{2005ApJ...635..386W}
{Wang}, Q.~D., {Yao}, Y., {Tripp}, T.~M., {et~al.} 2005, \apj, 635, 386

\bibitem[{{Willingale} {et~al.}(2013){Willingale}, {Starling}, {Beardmore}, {Tanvir}, \& {O'Brien}}]{Willingale.ea:13}
{Willingale}, R., {Starling}, R.~L.~C., {Beardmore}, A.~P., {Tanvir}, N.~R., \& {O'Brien}, P.~T. 2013, \mnras, 431, 394

\bibitem[{{Xu} {et~al.}(2023){Xu}, {Hao}, {Liu}, {Lin}, {Bian}, {Hou}, {Li}, \& {Li}}]{2023ApJ...947...54X}
{Xu}, Y., {Hao}, C.~J., {Liu}, D.~J., {et~al.} 2023, \apj, 947, 54

\bibitem[{{Zheng} {et~al.}(2024){Zheng}, {Ponti}, {Freyberg}, {Sanders}, {Locatelli}, {Merloni}, {Sasaki}, {Strong}, {Kerp}, {Maitra}, {Liu}, {Predehl}, {Anastasopoulou}, \& {Lamer}}]{Zheng2024}
{Zheng}, X., {Ponti}, G., {Freyberg}, M., {et~al.} 2024, \aap, 681, A77

\end{thebibliography}



\begin{appendix}

\section{Additional figures}
\begin{figure}
    \centering
    \includegraphics[width=0.49\textwidth]{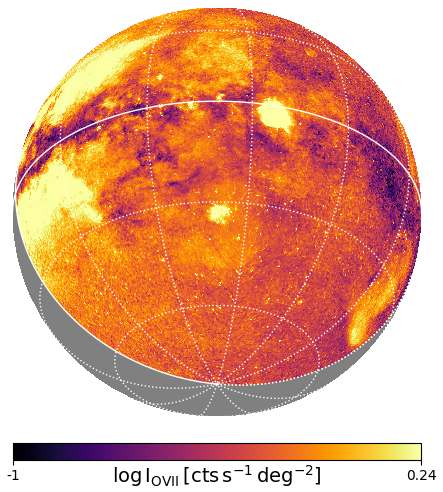}
    \includegraphics[width=0.49\textwidth]{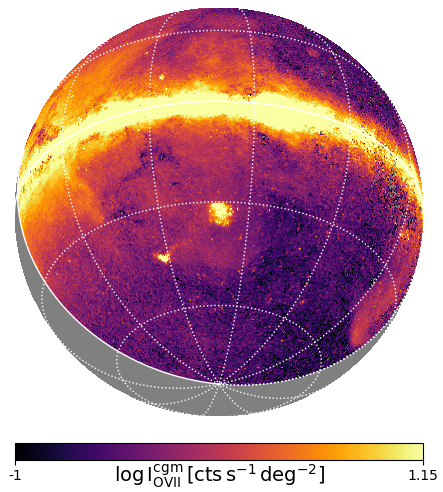}
    \caption{Same as Fig.~\ref{fig:hmap_O8cgm_n128} but for the \os{} narrowband (0.534-0.614 keV). See also Zheng et al. (submitted) for further details.}
    \label{fig:hmap_O7cgm_n128}
\end{figure}
\begin{figure}
    \centering
    \includegraphics[width=0.49\textwidth]{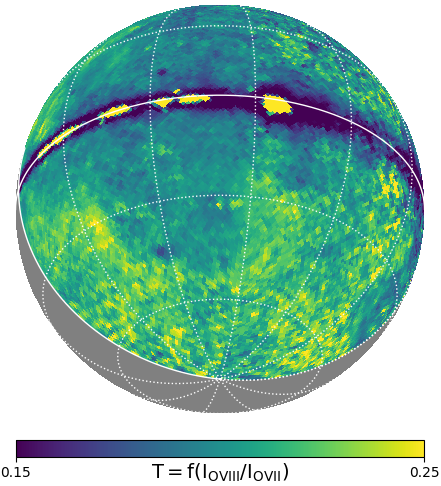}
    \caption{\erosita{} temperature proxy maps. \oo/\os{} (left panel) 
    Contributions from the LHB foreground, the instrumental background, the CXB and the foreground absorption have been removed from the map before computing the ratio. See the text and Zheng et al. (submitted) for further details. }
    \label{fig:temperature_maps}
\end{figure}
\begin{figure}
    \centering
    \includegraphics[width=0.49\textwidth]{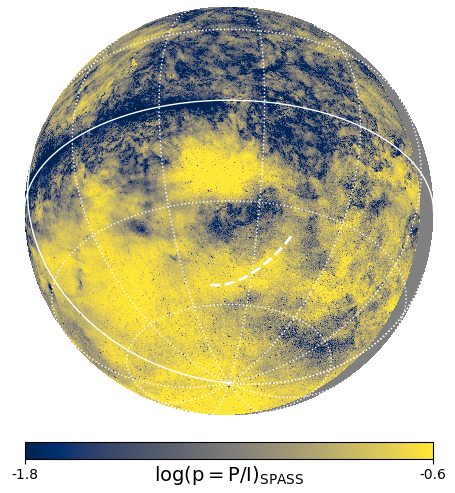}
    \caption{Same as Fig.~\ref{fig:spass_P} but for the fractional polarization $p=P/I$ of S-PASS \citep{2019MNRAS.489.2330C}.}
    \label{fig:spass_p}
\end{figure}

\end{appendix}


\label{lastpage}
\end{document}